\documentclass[twocolumn,showpacs,preprintnumbers,amsmath,amssymb,floatfix]{revtex4}

\usepackage{graphicx}
\usepackage{dcolumn}
\usepackage{bm}
\usepackage{color}
\usepackage{epstopdf}
\usepackage{here}

\newcount\timehh  \newcount\timemm
\timehh=\time \divide\timehh by 60 \timemm=\time
\definecolor{gold}{rgb}{0.85,.66,0}

\begin{document}

\title{Longitudinal and transversal spin dynamics of donor-bound electrons
in fluorine-doped ZnSe: spin inertia versus Hanle effect}

\author{F.~Heisterkamp$^{1}$, E.~A.~Zhukov$^{1}$, A.~Greilich$^{1}$,
D.~R.~Yakovlev$^{1,2}$, V.~L.~Korenev$^{1,2}$, A.~Pawlis$^{3,4}$ and M.
Bayer$^{1,2}$} \affiliation{$^1$Experimentelle Physik 2, Technische
Universit\"at Dortmund, 44221 Dortmund, Germany} \affiliation{$^2$Ioffe Physical-Technical Institute, Russian Academy of Sciences,
194021 St.~Petersburg, Russia} \affiliation{$^3$Peter Gr\"unberg Institute, Forschungszentrum J\"ulich, 52425 J\"ulich, Germany}
\affiliation{$^{4}$Department Physik, Universit\"at Paderborn, 33098 Paderborn, Germany}

\date{\today}

\begin{abstract}
The spin dynamics of the strongly localized, donor-bound electrons
in fluorine-doped ZnSe epilayers is studied by pump-probe Kerr
rotation techniques. A method exploiting the spin inertia is
developed and used to measure the longitudinal spin relaxation time,
$T_1$, in a wide range of magnetic fields, temperatures, and pump
densities. The $T_1$ time of the donor-bound electron spin of about
1.6~$\mu$s remains nearly constant for external magnetic fields
varied from zero up to 2.5~T (Faraday geometry) and in a temperature
range $1.8-45$~K. The inhomogeneous spin dephasing time,
$T_2^*=8-33$~ns, is measured using the resonant spin amplification
and Hanle effects under pulsed and steady-state pumping,
respectively. These findings impose severe restrictions on possible
spin relaxation mechanisms.
\end{abstract}


\maketitle

\section{Introduction}
\label{sec:intro}

Recently the coherent dynamics of spin excitations in semiconductor
heterostructures has attracted considerable interest, motivated in
particular by the observation of long electron spin relaxation and
coherence times~\cite{Awschalom_Spintronics}, one of the main
prerequisites for a system to be suited for quantum information
technologies. To obtain access to these times, optical techniques
have proofed to be an effective measurement tool.

Generally, the phenomenon of optical orientation is used to create
the initial spin orientation~\cite{OptOr}. It involves two
processes: the photogeneration of spin-oriented carriers by
absorption of circularly polarized light and the possible spin
relaxation with the characteristic time $\tau_{S}$ during the
lifetime $\tau$ of these carriers~\cite{OptOr}. In order to
determine absolute values of these times one often uses an ``internal
clock'' of the system: The periodic Larmor precession of the electron spins
about an external magnetic field with the frequency
$\Omega_{\text{L}}=\mu_{\text{B}}g_{\text{e}}B/\hbar$ can be used as
such a clock. Here $\mu_{\text{B}}$  is the Bohr magneton and
$g_{\text{e}}$ is the Land\'e factor of the electrons. One of the
common methods to study spin lifetimes
$T_{S}=1/(1/\tau+1/\tau_{S})$  in atoms ~\cite{Hanle1924,
Cohen} and in solid state systems ~\cite{Parsons1969, OptOr} is the
measurement of the Hanle effect. The Hanle effect analyzes the
decrease of the carrier spin polarization (typically via the
circular polarization degree of photoluminescence) in a transverse
magnetic field so that it also employs the clock defined by the
Larmor precession. For relatively strong magnetic fields, for which
the spin lifetime $T_{S}$  is long compared to the time scale
determined by the Larmor precession frequency $\Omega_{L}$
($T_{S}>>1/\Omega_{L}$), the electron spins perform many revolutions
during their lifetime~\cite{OptOr}. Thus, the spin polarization
along the direction of observation decreases with increasing
transverse magnetic field. The Hanle curve describes this behavior.
Its half-width at half maximum is given by
$B_{1/2}=\hbar/({\mu_{\text{B}}}g_{\text{e}}T_{S}$), so the spin
lifetime $T_{S}$ can be obtained by measuring the Hanle curve, if
the $g$ factor is
known~\cite{HanleReview1,HanleReview2,HanleReview3}.

Any Hanle effect-based method is based on the relaxation time
approximation, in which the dynamics is described by one or a few
exponents. It is a fair approximation, if the
relaxation is caused, for example, by processes with short correlation times
(Markovian processes) since these short correlation times lead to
dynamic averaging over magnetic fields of different origin, acting
on the electron. However, this approximation is violated for
strongly localized electrons, when the dwell time of the electron on
the donor exceeds the precession period of the electron spin in the
hyperfine field of the nuclei. The width of the Hanle curve for
donor-bound electron spins is determined by the relatively rapid
precession in static nuclear fields~\cite{Dzhi2002}, i.e. by the
spin dephasing time $T^*_2$, and not by the longitudinal spin
relaxation time $T_1$, which can be much longer than the precession
period in the frozen nuclear field. Precession in static fields is
reversible and to eliminate their effect the spin-echo method can be
used~\cite{Abragam}. However, this leads to a complication of
experiments on the irreversible spin dynamics, designed to determine
the $T_1$ time.

We propose a different approach to measure the spin lifetime, which
does not rely on the precession of the spins in a magnetic field
applied in the Voigt geometry. This method uses an external clock
instead of an internal one, namely the periodic polarization
modulation of the exciting light with the modulation frequency
$f_{\text{m}}$, and exploits the inertia of the spins: When
switching the helicity of the light the steady-state value of the
electron spin polarization is reached within a characteristic time
$T_{S}$. At low modulation frequencies
$2{\pi}f_{\text{m}}\ll1/T_{S}$, compared to the time scale given by
the spin lifetime, the electron spin polarization can overcome the
spin inertia and reach its steady-state value for a particular laser
polarization period. For high modulation
frequencies $2{\pi}f_{\text{m}}\ge1/T_S$, on the other hand, the
electron spin polarization remains reduced since it cannot reach its
steady-state value within a duty cycle with fixed circular
polarization. The fall or rise of the spin polarization in
dependence on the modulation frequency corresponds to the spin
lifetime. With this method one can measure the spin lifetime in a
weak magnetic field, when the dynamics of the average spin is
determined by relaxation processes in random fields that are not subject to
dynamic averaging, i.~e., when the method based on the Hanle effect
cannot provide the time $T_{S}$.

A similar method was used by Akimov et al.
\cite{I_Akimov,HennebergerSCQBs} to study the electron spin dynamics
in epitaxial CdSe/ZnSe quantum dots. The method combines time- and
polarization-resolved measurements of the emission from the trion
singlet ground state with helicity modulation of the exciting light.
However, the spin polarization was not measured in dependence of the
modulation frequency by Akimov et al., so our method can be seen as
an advancement. Fras et al. performed differential transmission
measurements of InAs/GaAs quantum dots using the optical pump-probe
technique~\cite{F_Fras}. Here, in addition to time-resolved
measurements a technique called dark-bright time scanning
spectroscopy was used, where the intensity of the exciting beam was
modulated to measure in the frequency domain.

Fluorine doped ZnSe recently emerged as a promising material system
in the field of solid-state quantum information technologies. So far
indistinguishable single-photon-sources and optically controllable
electron spin qubits were demonstrated \cite{Sanaka2012,
Sleiter2013, Kim2014}. Current efforts focus on gaining a detailed
understanding of the electron and nuclear spin dynamics in this
material.

Here, using the spin inertia method we investigate the spin dynamics
of the strongly localized, donor-bound electrons in fluorine-doped
ZnSe epilayers in a wide range of magnetic fields, temperatures, and
pump densities. The paper is organized as follows.
Section~\ref{sec:1} provides details of the experimental techniques
and studied samples.  Section~\ref{sec:2} describes the experimental
results. Section~\ref{sec:theory} is devoted to the theoretical
consideration of the spin inertia effect. Modeling of the experimental
data can be found in Section~\ref{sec:modeling}. Discussion of the
spin relaxation mechanisms is done in Section~\ref{sec:discussion}
in combination with assessments on the applicability of the spin
inertia method to various spin systems.

\section{Experimental Details}
\label{sec:1}

We study two fluorine-doped ZnSe epilayers with different doping
concentrations. The samples consist of three layers grown by
molecular-beam epitaxy on (001)-oriented GaAs substrate. A thin ZnSe
buffer layer reduces the strain induced by the II-VI on III-V
heteroepitaxy. The ZnSe layer is followed by a 20-nm-thick
Zn$_{1-x}$Mg$_{x}\text{Se},~x<0.15$ barrier layer, which prevents
carrier diffusion into the substrate. The fluorine-doped,
70-nm-thick ZnSe epilayer is grown on top of this barrier layer.
Sample \#1 has a fluorine concentration of about $1\times10^{15}$
cm$^{-3}$, while the doping of sample~\#2 is approximately three
orders of magnitude higher ($1\times10^{18}$ cm$^{-3}$). For the
optical properties of these samples and for information on the
electron spin dephasing we refer to Ref.~\cite{Greilich12}.

The samples are placed in a vector magnet system consisting of three
superconducting split-coils oriented orthogonally to each other
\cite{Zhukov2012}. It allows us to switch the magnetic field from the
Faraday geometry (magnetic field $B_{\text{F}}$ parallel to the
sample growth axis and the light wave vector) to the Voigt geometry
(magnetic field $B_{\text{V}}$ perpendicular to the sample growth
axis and the light wave vector). The switching can be performed by
using the respective pairs of split coils and does not require any
changes of the optical alignment. Therefore we can measure in
different magnetic field geometries with exactly the same adjustment
of the pump and probe beams on a particular sample position. The
measurements are performed at low temperatures with the samples
either immersed in pumped liquid helium at $T=1.8$~K or cooled with
controlled helium gas flow (up to 45 K). Photoluminescence (PL)
spectra for sample characterization are excited using a
continuous-wave (CW) laser with photon energy of 3.05~eV and
detected with a Si-based charge-coupled device (CCD) camera attached
to a 0.5-m spectrometer.

We use the pump-probe technique to study the electron spin dynamics
by time-resolved Kerr rotation (TRKR). The electron spin coherence
is created by circularly-polarized pump pulses of 1.5~ps duration
(spectral width of about 1~meV) emitted by a mode-locked Ti:Sapphire
laser operating at a repetition frequency of 75.7~MHz (repetition
period $T_{\text{R}}=13.2$~ns). The induced electron spin coherence
is measured by linearly-polarized probe pulses of the same photon
energy as the pump pulses (degenerate pump-probe scheme). A
mechanical delay line is used to scan the time delay between the
probe and pump pulses. The photon energy is tuned into resonance
with the donor-bound heavy-hole exciton (D$^{0}$X-HH) at about
2.80~eV. To obtain this photon energy a Beta-Barium Borate (BBO)
crystal is used to double the frequency of the light generated by
the Ti:Sapphire laser. The pump helicity is modulated between
$\sigma^+$ and $\sigma^-$ polarization by an electro-optical
modulator (EOM), so that on average the samples are equally exposed
to left- and right-circularly-polarized pump pulses. The modulation
frequency is varied between 10~kHz and 700~kHz. The photogenerated
spin polarization results in a rotation of the polarization plane of
the reflected, initially linear-polarized probe pulses due to the
magneto-optical Kerr effect. The Kerr rotation (KR) angle is
measured by a 10~MHz balanced photoreceiver with adjustable gain and
bandwidth, connected to a lock-in amplifier. The pump density is
varied in the range $P_{\mathrm{pump}}=0.2-4.2$~W/cm$^2$ and the
probe density ($P_{\mathrm{probe}}$) is about one order of magnitude
smaller than the pump density.

We use three different implementations of the pump-probe Kerr
rotation method:

(1) The time-resolved Kerr rotation configuration, where the Kerr
rotation angle is measured in dependence of the time delay between
the pump and probe pulses with the magnetic field applied in the
Voigt geometry. In this case the Larmor precession of the electron
spin polarization around the magnetic field axis results in a signal
which is a periodic function of the time delay and whose amplitude
decreases with increasing time delay. Using this configuration one
can determine the $g$ factor of the carriers and the inhomogeneous
spin dephasing times $T_{2}^{*}$ in the limit
$T_{2}^{*}<T_{\text{R}}$~\cite{Greilich12}.

(2) The resonant spin amplification (RSA)
configuration~\cite{Kikkawa98,Awschalom_Spintronics,Yugova12} is
used to determine $T_{2}^{*}$ when this time is comparable to or
greater than the laser repetition period $T_{\text{R}}$. Here the
time delay between pump and probe is fixed at a small negative value
($\Delta t \approx-20$~ps) and one measures the KR angle in
dependence of the magnetic field applied in the Voigt geometry in
the range from $-20$ to $+20$~mT. At certain magnetic fields the
electrons spins precess in phase with the laser repetition frequency
and one observes an increased Kerr rotation signal. Thus, the RSA
signal consists of a symmetrical set of equidistant peaks, whose
amplitude decreases with increasing magnetic field.

(3) In the polarization recovery (PR) configuration the electron
spin polarization is detected as well at a small negative time
delay. The KR signal is measured in dependence of the magnetic field
applied in the Faraday geometry. The electron spin polarization,
which is photogenerated along the magnetic field direction, does not
exhibit Larmor precession then. Still it decreases by the nuclear
hyperfine fields, if the external magnetic field is small compared
to these fields. The effect of the hyperfine fields is suppressed
with increasing external magnetic field. As a result, the
electron spin polarization has its minimum at zero external magnetic
field and increases with magnetic field. By varying the pump
helicity modulation frequency one can measure the longitudinal spin
relaxation time $T_1$ of the electrons. We will mostly use this
implementation to study the spin dynamics of the donor-bound
electrons.

Note that the measurement of the KR signals at negative time delay,
prior to the pump pulse, as done in the RSA and PR configurations
greatly simplifies the interpretation of the signal origin. These
signals can only arise from long-living spins, whose lifetime
exceeds $T_\mathrm{R}=13.2$~ns. This is typically much longer than
the exciton recombination time, so that the measured signals can
originate only from resident electrons, which are bound to donors at
low temperatures.

In addition, we also perform pump-probe experiments using a CW pump
and a pulsed probe. For these measurements a CW Ti:Sapphire laser
with intra-cavity second harmonic generation is used as the pump,
and the probe pulses are generated from the laser system described
above. This configuration allows us to set the pump and the probe
laser at different photon energies, i.e. to perform two-color
nondegenerate pump-probe measurements. Thereby we measure the PR and
the suppression of the KR signal in the Voigt geometry (the Hanle
curve), to investigate possible influences of pulsed excitation on
the spin relaxation.

\section{Experimental results}
\label{sec:2}

\begin{figure}[bt]
\includegraphics[width=1.0\linewidth]{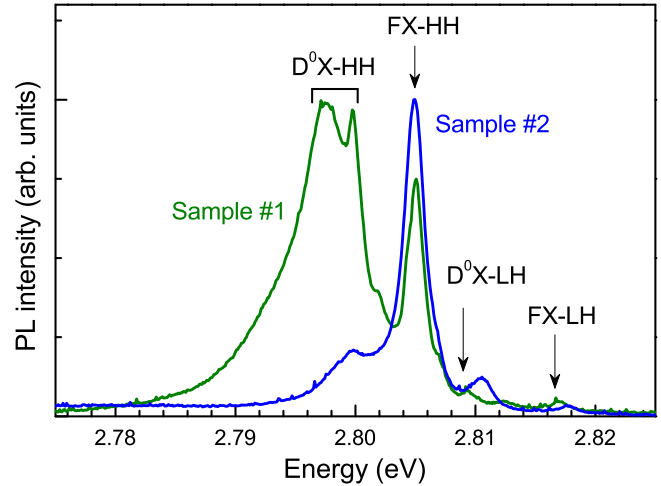}
\caption[] {(Color online) PL spectra of the fluorine-doped ZnSe
epilayers \#1 and \#2  measured at $B=0~$T for $T=1.8~$K.}
\label{fig:1}
\end{figure}

Figure~\ref{fig:1} shows the PL spectra of the two studied samples,
measured at zero magnetic field for a temperature of $T=1.8~$K. The
spectrum of sample \#1 exhibits the following emission lines:
Donor-bound heavy-hole exciton (D$^{0}$X-HH) at $2.7970-2.7997$~eV,
free heavy-hole exciton (FX-HH) at $2.8045$~eV, donor-bound
light-hole exciton (D$^{0}$X-LH) at $2.8092$~eV and free light-hole
exciton (FX-LH) at $2.8167$~eV~\cite{Greilich12}. The strain induced
by the II-VI on III-V heteroepitaxy lifts the light-hole and
heavy-hole degeneracy for both structures~\cite{Greilich12}. The
donor-related lines for the higher doped sample \#2 exhibit a small
blue-shift, compared to the same lines for sample \#1.

\begin{figure}[bt]
\includegraphics[width=1.0\linewidth]{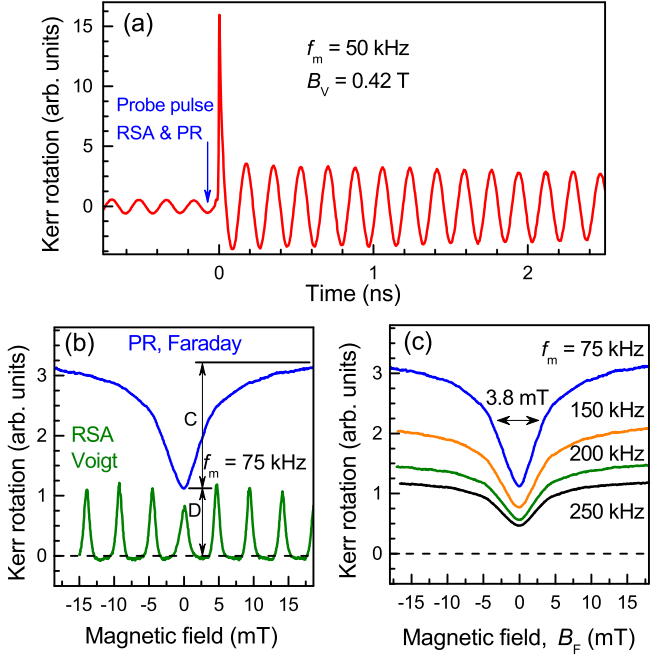}
\caption[] {(Color online) TRKR results for sample~\#1 measured for
resonant D$^{0}$X-HH excitation (2.7986~eV) at $T=1.8$~K. (a) KR
signal in dependence on the time delay at $f_\mathrm{m} = 50$~kHz
and $P_{\text{pump}}=1.6$~W/cm$^{2}$.  The arrow marks the time
delay at which the RSA and PR signals are detected. (b) PR and RSA
signals measured at $f_\mathrm{m} = 75$~kHz. (c) PR signals measured
at different modulation frequencies. In panels (b) and (c)
$P_{\text{pump}}=0.5$~W/cm$^{2}$.} \label{fig:2}
\end{figure}

Results of pump-probe measurements in all three experimental
configurations are illustrated in Fig.~\ref{fig:2} for sample~\#1.
Results obtained with the TRKR and RSA configurations were
considered in detail in Ref.~\cite{Greilich12} and are given here
for illustration and comparison with the PR data. Furthermore, they
provide important supplementary information on the donor-bound
electron spins.

Figure~\ref{fig:2}(a) shows a time-resolved Kerr rotation signal
measured at a temperature of $T=1.8$~K for resonant D$^{0}$X-HH
excitation. A magnetic field of $B_{\text{V}}=0.42$~T is applied in
the Voigt geometry and the observed oscillations reflect the Larmor
precession of the electron spin polarization. Note that these
oscillations are long-living and do not fully decay during the time
interval $T_\mathrm{R}=13.2$~ns between subsequent pump pulses, as
can be seen from the considerable signal amplitude at negative time
delays. The exciton lifetime in ZnSe is shorter than 250~ps
\cite{Greilich12}, which allows us to conclude that the long-living
TRKR signal originates from the coherent spin precession of the
localized donor-bound electrons. The relatively large binding energy
to these donors of 29~meV \cite{Merz72} provides strong electron
localization and makes the spin coherence robust even at elevated
temperatures up to 40~K \cite{Greilich12}. We evaluate a $g$ factor
of the donor-bound electron of $|g_{\text{e}}|=1.13\pm0.02$ from the
period of the signal oscillations. The same value of
$|g_{\text{e}}|$ is measured for sample~\#2.

Due to the long decay of the TRKR signal amplitude it is difficult
to evaluate the electron spin dephasing time $T_{2}^{*}$ by fitting
the amplitude decay in these measurements. Instead we use the RSA
technique for that purpose, for details see Ref.~\cite{Greilich12}.
An example of a RSA signal is shown by the green line in
Fig.~\ref{fig:2}(b). The analysis of the RSA peak width at $T=1.8$~K
for low pump densities yields $T_{2}^{*}=33$~ns for the sample~\#1
and 8~ns for the sample~\#2. The faster dephasing in the sample with
higher fluorine concentration can be explained by the interaction
between electron spins localized at neighboring
donors~\cite{Greilich12}.

The polarization recovery signal measured for the same experimental
conditions as the RSA signal (only the magnetic field geometry is
changed from Voigt to Faraday) is shown in Fig.~\ref{fig:2}(b) by
the blue line. The PR curve has a minimum at zero magnetic field,
increases with increasing $B_{\text{F}}$ and saturates at fields
exceeding 20~mT. Obviously, the polarization recovery is caused by
suppression of the depolarization of the electron spin along the
magnetic field direction. We tentatively relate the depolarization
around zero field to the effect of the fluctuating nuclear hyperfine
fields, more details will be given in the discussion below.

As shown in Figure~\ref{fig:2}(b), the amplitude of the zero RSA
peak is a little smaller than the amplitude of the neighboring
peaks. This may be due to the following factors: 1) A small,
additional magnetic field component perpendicular to $B_{\text{V}}$
can lead to a reduction of the zero RSA peak amplitude
\cite{Zhukov2012}. This component can occur if there is a small
inclination (about $1-2^\circ$) of the sample plane with respect to
the $\mathbf{k}$-vector (either horizontally or vertically). 2) An
additional nuclear field induced at $B_{\text{V}}$ may also lead to
a a reduction or an increase of the amplitude of the zero RSA peak
\cite{Zhukov2014}.

Figure~\ref{fig:2}(c) shows PR signals, measured for different pump
helicity modulation frequencies, $f_{\text{m}}$, varied from 75 up
to 250~kHz. The magnitude of the PR signal decreases for higher
$f_{\text{m}}$, while the full width at half maximum (FWHM) of the
dip around the zero magnetic field of about 3.8~mT and the overall shape
of the PR curves remain the same. From these findings one can
suggest that the inverse electron spin relaxation time falls in the
examined frequency range.

\begin{figure}[bt]
\includegraphics[width=1.0\linewidth]{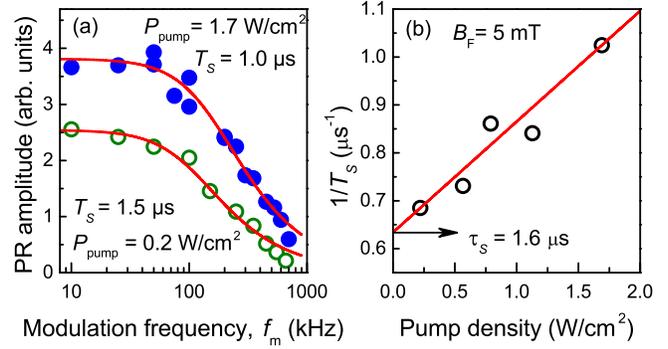}
\caption[] {(Color online) Spin dynamics measured for sample~\#1
with the PR technique at $B_{\text{F}}=5$~mT for $T=1.8$~K. (a) PR
amplitude in dependence of $f_{\text{m}}$ for two pump densities of
0.2 and 1.7~W/cm$^{2}$. Red lines show the fits to the data based on
our theoretical model (cf. Eq.~\eqref{n8}), which is used to
determine the spin lifetime $T_S$. (b) Inverse spin lifetime $1/T_S$
in dependence of the pump density. Red line is linear fit to the
data, which is used to extrapolate the spin relaxation time
$\tau_S=1.6~\mu$s.} \label{fig:3}
\end{figure}

To study this in more detail the PR amplitude in dependence of the
modulation frequency is measured at $B_{\text{F}}=5$~mT for two pump
densities.  The PR amplitude for both pump densities, shown by the
symbols in Fig.~\ref{fig:3}(a), remains constant for low modulation
frequencies on the order of a few 10~kHz, while it rapidly decreases
above 100~kHz. Model calculations shown by the red lines (details
will be given in  Secs.~\ref{sec:theory} and \ref{sec:modeling})
allow us to evaluate the spin lifetime $T_S=1.5~\mu$s for
$P_{\text{pump}}=0.2$~W/cm$^{2}$ and $T_{S}=1.0~\mu\text{s}$ for
$P_{\text{pump}}=1.7$~W/cm$^{2}$. The spin lifetime in dependence of
the pump density is plotted in Fig.~\ref{fig:3}(b). The decrease of
the PR amplitude with increasing $f_{\text{m}}$ is the key result of
this study. In the following we present details of its change with
varying magnetic field strength and temperature in order to obtain
comprehensive information on the spin dynamics of the donor-bound
electrons in ZnSe.

\begin{figure}[bt]
\includegraphics[width=1.0\linewidth]{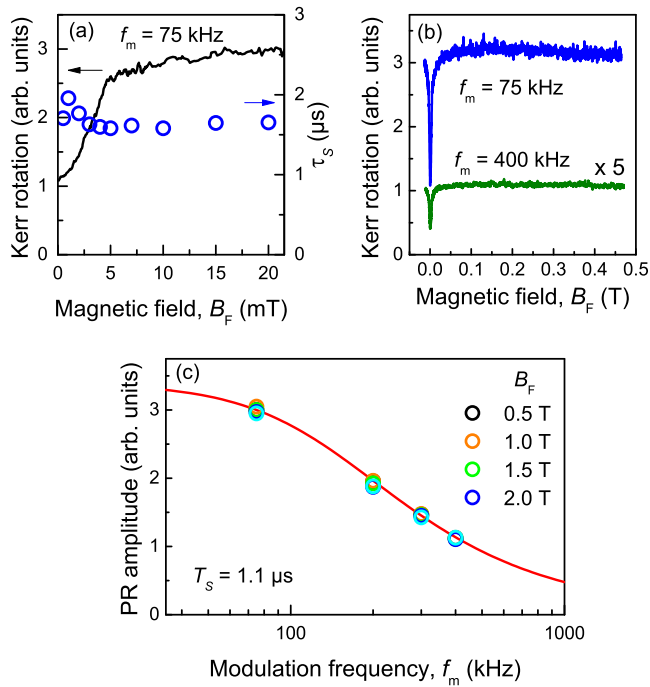}
\caption[] {(Color online) Results for sample~\#1 measured at
$T=1.8$~K. (a) Blue circles give the spin relaxation time $\tau_{S}$
in dependence of the magnetic field. Black line shows a typical PR
signal. $P_{\text{pump}}=1.7$~W/cm$^{2}$. (b) PR signals in
dependence of $B_{\text{F}}$ for $f_{\text{m}}=75$~kHz (blue line)
and $400~$kHz (green line). $P_{\text{pump}}=2.4$~W/cm$^{2}$. (c)
Modulation frequency dependence of the PR amplitude measured in
different magnetic fields $B_{\text{F}}$.
$P_{\text{pump}}=2.4$~W/cm$^{2}$. Red line shows fit to the data
according to Eq.~\eqref{n8} with the fit parameter $T_S=1.1~\mu$s.}
\label{fig:4}
\end{figure}

The blue circles in Fig.~\ref{fig:4}(a) illustrate the spin
relaxation time $\tau_{S}$, determined with our model, in dependence
of the magnetic field, varied from zero to 20~mT. The black line
shows the corresponding PR signal at $f_{\text{m}}=75~\text{kHz}$.
The spin relaxation time remains constant within the accuracy of our
method in this $B_{\text{F}}$ range. The PR signal in an extended
magnetic field range up to 0.5~T is shown in Fig.~\ref{fig:4}(b).
One sees that the signal is pretty much constant in the field range
from 0.02 to 0.5~T. In this range its amplitude decreases by a
factor of 15, when $f_{\text{m}}$ is changed from 75~kHz to 400~kHz
(note the multiplication factor of 5 in the figure). For higher
fields we perform measurements each 0.5~T in the range 1.0-2.5~T.
For each field four modulation frequencies are examined (see
Fig.~\ref{fig:4}(c)). For all measured $f_{\text{m}}$ the PR
amplitude is independent of the magnetic field strength. Its
frequency dependence can be fitted with the same function shown by
red line. From this fit we obtain $T_S=1.1~\mu$s. An important
experimental result of Fig.~\ref{fig:4} is that the PR amplitude in
dependence of $B_{\text{F}}$ considerably increases from zero to
20~mT, but then remains constant in the range from 20~mT up to
2.5~T.

\begin{figure}[bt]
\includegraphics[width=7 cm]{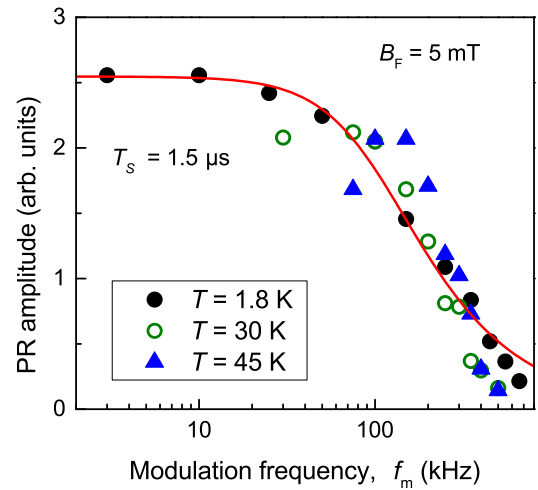}
\caption[] {(Color online) PR amplitude in dependence of the
modulation frequency measured for sample~\#1 at three different
temperatures. Data are normalized to each other at
$f_{\text{m}}=100$~KHz. $P_{\text{pump}}=0.4$~W/cm$^{2}$. Red line
shows fit to the data at $T=1.8$~K according to Eq.~\eqref{n8} with
the fit parameter $T_S=1.5~\mu$s.} \label{fig:5}
\end{figure}

The shape of the PR amplitude as function of the modulation
frequency is maintained in the temperature range from 1.8 up to
45~K, as illustrated by the experimental data presented in
Fig.~\ref{fig:5}, where results for $T=1.8$, 30 and 45~K are
compared. The PR amplitude decreases slightly by less than 40\% for
elevated temperatures and has been normalized to $T=1.8~\text{K}$ at
100~kHz. The shape of the frequency dependence remains almost the
same evidencing that the spin dynamics of the donor-bound electrons
does not change at $T<45$~K.

The spin relaxation mechanism of optically oriented carriers may
depend on whether CW or pulsed photoexcitation is used. Excitation
of spins systems with short pulses of picosecond duration may induce
perturbations assisting the spin relaxation. This problem has been
addressed in Ref.~\cite{Astakhov2008}, where the coherent spin
dynamics of resident electrons in n-doped CdTe/(Cd,Mg)Te quantum
wells has been compared for CW and pulsed excitation. But in this
case the in-plane localizing potential of about 1~meV for the
resident electrons formed by monolayer well width fluctuations is
weak compared with the fluorine-donor binding energy of 29~meV in
ZnSe.

\begin{figure}[hbt]
\includegraphics[width=8 cm]{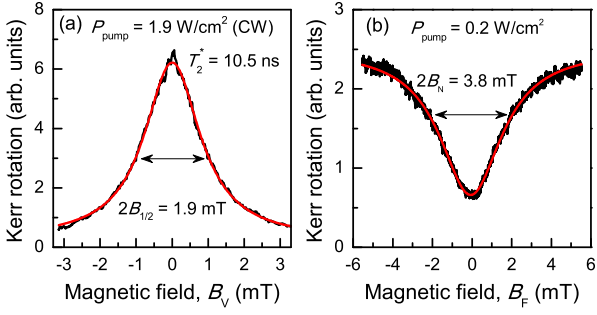}
\caption[] {(Color online) Spin dynamics of sample~\#2 measured by
two different techniques at $T=1.8$~K and $f_{\text{m}}=50$~KHz. (a)
Hanle curve induced by a CW pump resonant with the D$^{0}$X-LH
transition (2.8092~eV) and detected with a  pulsed probe at the
D$^{0}$X-HH transition (2.7986~eV). Red line shows fit with a
Lorentz curve (see Eq. \eqref{n12} in Sec. \ref{sec:modeling})
giving $B_{1/2}=0.95$~mT corresponding to $T^*_2=10.5$~ns. (b) PR
curve measured with pulsed pump and probe beams, both resonant with
the D$^{0}$X-HH transition (2.7986~eV). Red line shows fit to the
data using Eq.~\eqref{e8a} with the fit parameter
$B_\mathrm{N}=1.9$~mT.} \label{fig:6}
\end{figure}

The spin dynamics of the donor-bound electrons under CW pump
measured for sample~\#2 is shown in Fig.~\ref{fig:6}(a). In this
experiment the pump and the probe have different photon energies.
The CW pump is resonant with the D$^{0}$X-LH transition and the
pulsed probe detects the electron polarization at the D$^{0}$X-HH
transition. Typically the energy relaxation between light-hole and
heavy-hole exciton states is fast and does not lead to considerable
losses in optical orientation. The magnetic field is applied in the
Voigt geometry. A decrease of the spin polarization with increasing
magnetic field is observed. This depolarization with increasing
field can be assigned to the Hanle effect. The width of the Hanle
curve is $2B_{1/2}=1.9$~mT. For comparison in Fig.~\ref{fig:6}(b) a
PR curve for sample~\#2 is shown. It has a width of 3.8~mT, which is
similar to the results for sample~\#1 (see Fig.~\ref{fig:2}(c)).

In the next two sections the chosen theoretical approach for
describing the PR effect and its dependence on the pump helicity
modulation frequency and the pump density is developed. The
experimental data are fitted based on this model and the
characteristic times for the electron spin dynamics are evaluated.

\section{Theory}
\label{sec:theory}

In n-type semiconductors the process of optical orientation results
from the replacement of unpolarized resident electrons by
photogenerated spin-oriented electrons \cite{OptOr}. The electrons
loose their spin orientation due to spin relaxation with time
$\tau_{S}$. Also their recombination with photogenerated holes will
reduce the macroscopic electron spin polarization. As a result, the
lifetime of the photogenerated electrons, $\tau=n_{0}/G$, depends on
the rate of electron-hole generation $G$, and on the resident
electron concentration $n_{0}$. The spin lifetime $T_S$
\begin{equation}
\label{n0.1}
1/T_{S}=1/{\tau}+1/{\tau_{S}}
\end{equation}
determines the time until the steady-state spin polarization is
reached by optical pumping.

\begin{figure}[bt]
\includegraphics[width=1.0\linewidth]{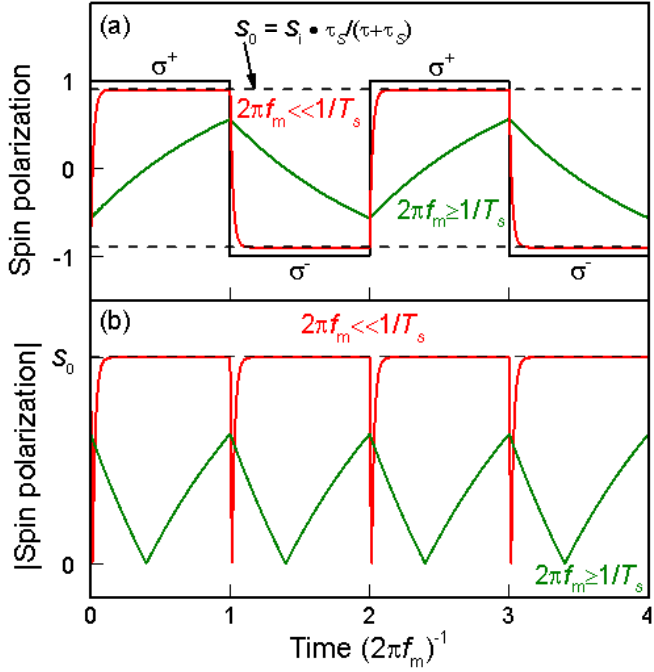}
\caption[] {(Color online) Illustration of the effect of electron
spin inertia for pump helicity modulation with frequency
$f_{\text{m}}$. The red and green lines show the limits of $2\pi
f_{\text{m}}\ll 1/T_{S}$ and $2\pi f_{\text{m}}\ge 1/T_{S}$,
respectively. (a) Illustration of the spin polarization $S_{z}$
along the direction of observation for two modulation periods: While
in the first case the spin polarization follows the laser
polarization without inertia and always reaches the steady state
value $S_{0}$ in a fixed laser polarization period, the spin
polarization cannot reach $S_0$ during such a period, when the
modulation is fast compared to the time scale given by the spin
lifetime $T_{S}$. (b) Modulus of the spin polarization $|S_{z}(t)|$
for both limits. While in the first case ($2\pi f_{\text{m}}\ll
1/T_{S}$, red line) $|S_{z}|$ is equal to $\vert S_0\vert$ almost
during the whole modulation period except for a small decrease, when
the sign of the polarization is switched, the modulus of the spin
polarization is strongly decreased during the whole modulation
period for $2\pi f_{\text{m}}\ge 1/T_{S}$.} \label{fig:7}
\end{figure}

In our experiment in the polarization recovery configuration the
pump helicity is modulated, so that the spin polarization is
switched between steady-state polarizations with opposite signs. On
the one hand, if the modulation frequency is so small that the
period with constant pump helicity is much longer than the spin
lifetime ($2\pi f_{\text{m}}\ll 1/T_{S}$), the average spin
polarization seems to follow the pump polarization with negligible
``inertia'' effects (see Fig.~\ref{fig:7}). On the other hand, if
the pump helicity modulation is so fast that the period with
constant pump helicity is comparable to or shorter than the spin
lifetime ($2\pi f_{\text{m}}\ge 1/T_{S}$), the spin polarization
cannot reach its steady-state value and the Kerr rotation signal is
decreased significantly.

Let us consider the case when the electron spin polarization
$\mathbf{S}$ is generated along the $z$-axis, i.e. when the light
wave vector of the pump laser is parallel to the $z$-axis
($\mathbf{k}\parallel \mathbf{z}$). The amplitude of the Kerr
rotation signal is proportional to $S_z$. The following kinetic
equation describes the dynamics of the electron spin
polarization~\cite{OptOr}:
\begin{equation}
\label{n1} \frac{dS_z(t)}{dt} =
\frac{S_{\text{i}}-S_z(t)}{\tau}-\frac{S_z(t)}{\tau_{S}} .
\end{equation}
The initially generated spin polarization
$\textbf{S}_{\text{i}}=(0,0,S_\mathrm{i})$ depends on the laser
polarization and optical selection rules. The first term in the
right equation part describes the polarization injection
($S_\text{{i}}/\tau$) and escape due to electron recombination
($-S_z/\tau$) with time $\tau$, and the second term describes the
spin relaxation with time $\tau_{S}$.

For constant circular polarization of the pump the stationary solution is:
\begin{equation}
\label{n2} S_{z}=S_{0}=S_{\text{i}}\frac{\tau_{S}}{\tau_{S}+\tau} =
S_{\text{i}}\frac{G\tau_{S}}{G\tau_{S}+n_{0}}  .
\end{equation}
For pump helicity modulation with a frequency $f_{\text{m}}$ we have
to solve the non-stationary Eq.~\eqref{n1}. Combining Eqs.
\eqref{n0.1} and \eqref{n1}, we find
\begin{equation}
\label{n3} \frac{dS_{z}(t)}{dt}=\frac{S_{0}(t)-S_{z}(t)}{T_{S}}.
\end{equation}
In our experiment
$S_{0}(t)=S_{\text{i}}(t)\frac{\tau_{S}}{\tau+\tau_{S}}$ is an
alternating signal of rectangular pulses with a constant amplitude
$|S_{0}|$, a duty cycle of 0.5 and the modulation frequency
$f_{\text{m}}$.

In the PR experiment we measure the Kerr rotation signal, which is
proportional to $n_{0}S_{z}$. The spin polarization along the
direction of observation $S_{z}(t)$ is oscillating with the
modulation frequency  $f_{\text{m}}$. This means that the we measure
the following correlator:

\begin{eqnarray}
\label{n4}
L(f_{\text{m}})&=&{\langle}S_{z}(t)\exp(i2{\pi}t/T_{\text m}){\rangle}|_{T_{\text{m}}}=\nonumber\\
&&\int\limits_0^{T_{\text{m}}}\frac{S_{z}(t)\exp(i2{\pi}t/T_{\text{m}})}{T_{\text{m}}}dt .
\end{eqnarray}
The averaging is done over the pump modulation period
$T_{\text{m}}=1/(2\pi f_{\text{m}})$. As a result, the task consists
of two steps: (i) determine $S_{z}(t)$ and (ii) calculate the
correlator according to Eq. \eqref{n4}. The calculations show that
the spin polarization along the direction of observation,
$S_{z}(t)$, is a periodic function with the period $T_{\text{m}}$ of
the pump helicity modulation:
\begin{equation}
\label{n5}
S_{z}(t)=|S_{0}|\left(1-\frac{2\text{e}^{-\frac{t}{T_{S}}}}{1+\text{e}^{-\frac{T_{\text{m}}}{2T_{S}}}}\right)
,
\end{equation}
in the half cycles in which $S_{0}(t)=+|S_0|$.
\begin{equation}
\label{n6} S_{z}(t)=|S_{0}|\left\{-1+2\left(\text{e}^{
 \frac{T_{\text{m}}}{2T_{S}}}-\frac{1}{1+e^{\frac{-T_{\text{m}}}{2T_{S}}}}\right)\text{e}^{-\frac{1}{T_{S}}}\right\},
\end{equation}
in the half cycles in which $S_{0}(t)=-|S_0|$. Hence it is possible
to determine the following correlator \eqref{n4}:
\begin{equation}
L(f_{\text{m}})=-\frac{2n_0 \vert S_0 \vert}{\pi (i+2{\pi}f_{\text{m}}T_S)} .
\label{n7}
\end{equation}
In the experiment the lock-in amplifier records the following
signal:
\begin{equation}
|L(f_{\text{m}})|=\frac{2}{\pi}\frac{ n_0 \vert S_0\vert}{\sqrt{1+(2{\pi}f_{\text{m}}T_S)^2}}  .
\label{n8}
\end{equation}
We use this expression to fit the PR amplitude in dependence of the
modulation frequency, which allows us to evaluate $T_S$, as the only
fitting parameter, by the spin inertia effect.

Figure~\ref{fig:7} illustrates schematically the spin inertia effect
and shows how the dependence of the correlator on the modulation
frequency and the spin lifetime manifests itself in experiment.
Figure~\ref{fig:7}(a) shows the time-dependent spin polarization
along the direction of observation $S_{z}(t)$ for two modulation
periods. In the case of slow modulation compared to the spin
lifetime (red line) the spin polarization follows the laser
polarization without significant inertia and always reaches the
steady state value $\vert S_0 \vert$ in a period of fixed laser
polarization. However, the spin polarization cannot reach $ \vert
S_0 \vert$ during such a period, when the modulation occurs fast
compared to the spin lifetime $T_{S}$ (green line). Note that the
time-averaged spin polarization is equal to zero in both cases.
However, the lock-in amplifier records the signal, which is
proportional to the time-averaged modulus $|S_z|$ of the spin
polarization (see Fig.~\ref{fig:7}(b)). For $2\pi f_{\text{m}}\ll
1/T_{S}$ (red line) this time-averaged value is very close to $\vert
S_0 \vert$, while it is clearly smaller than $\vert S_0 \vert$ in
the limit $2\pi f_{\text{m}}\ge 1/T_{S}$. We denote this
impossibility of the spin polarization to follow the polarization of
the exciting light for fast modulation compared to the spin lifetime
as the spin inertia effect.

Investigation of the carriers spin dynamics by the spin inertia
effect can be performed at zero as well as finite external magnetic
fields. In magnetic field the evaluated spin relaxation time
$\tau_S$ corresponds to the longitudinal spin relaxation time $T_1$.
It is also valid for samples in which the electron spins are
affected by randomly oriented hyperfine fields from the nuclear spin
fluctuations~\cite{Dzhi2002,Merkulov2002}, namely, when the spin
dephasing time $T_{2}^{*}$ caused by the nuclear fluctuations is
considerably shorter than $T_1$. Note, that in this case the method
based on the Hanle effect is limited to measurements of the
$T_{2}^{*}$ time, and not the $T_1$ time, we discuss this in more
detail in Sec.~V.

For example, in our fluorine-doped ZnSe samples the donor-bound
electrons are strongly localized. Thus, the dwell time of an
electron on a donor is longer than the inhomogeneous dephasing time
$T_{2}^{*}$ of the ensemble of donor-bound electrons in the frozen
hyperfine fields of the nuclei, $\mathbf{B}_\mathrm{N}$. The
components of the electron spin perpendicular to the hyperfine field
decay during time $T_{2}^{*}$, while the spin polarization along the
hyperfine field direction decays on a much longer time scale,
$T_1\gg T_{2}^{*}$. The effect of the fluctuating hyperfine fields
can be taken into account in the model by adding the term
$\mathbf{\Omega}_\mathrm{N}\times\mathbf{S}$ to
Eq.~\eqref{n1}~\cite{Dzhi2002}:
\begin{equation}
\label{n1addnuclei} \frac{d\mathbf{S}(t)}{dt} =
\frac{\mathbf{S}_{\text{i}}-\mathbf{S}(t)}{\tau}-\frac{\mathbf{S}(t)}{\tau_{S}}
+\mathbf{\Omega}_\mathrm{N}\times\mathbf{S}.
\end{equation}
Here
$\mathbf{\Omega}_\mathrm{N}=\mu_{\text{B}}g_{\text{e}}\mathbf{B}_\mathrm{N}/\hbar
$. The term $\mathbf{\Omega}_\mathrm{N}\times\mathbf{S}$ describes
the precession of the spins in the frozen nuclear field with
subsequent averaging over the Gaussian distribution of these
fields~\cite{Dzhi2002}. Here one averages over an isotropic
distribution of nuclear fields. Thus, the generated electron spin
polarization decreases to one third of its initial
value~\cite{Dzhi2002,Merkulov2002}. Under these conditions the width
of the Hanle curve is determined by the shortest characteristic time
$T_{2}^{*}$~\cite{Dzhi2002}. However, the cutoff frequency on the
spin inertia effect is still determined by the spin lifetime
$T_{S}$, which originates from the $T_1$ time. Due to the hyperfine
fields the initial spin $S_{0}$ in Eq.~\eqref{n8} depends on the sum
of the external magnetic field and the hyperfine field averaged over
the realizations of randomly oriented nuclear fields (see
Ref.~\cite{Merkulov2002}).

\section{Modeling of experimental data}
\label{sec:modeling}

We interpret the decrease of the PR amplitude for increasing
modulation frequency, shown in Figs.~\ref{fig:2}(c)  and
\ref{fig:3}(a), as a decrease of the electron spin polarization due
to the spin inertia effect. The red lines in Fig.~\ref{fig:3}(a) are
fits to the data according to Eq.~\eqref{n8} for sample~\#1. From
these fits we obtain spin lifetimes of $T_{\text{S}}=1.5$ and
1.0$~\mu$s for low and high pump density, respectively. The pump
density dependence of $T_S$ is described by Eq.~\eqref{n0.1}.
Keeping in mind that $\tau=n_{0}/G$, one sees that for vanishing
pump rates $G$ the term $1/\tau \to 0$ and $T_S \to \tau_{S}$. This
gives us  a way to measure the spin relaxation time $\tau_{S}$.
Figure~\ref{fig:3}(b) shows the inverse spin lifetime $1/T_{S}$ in
dependence of the pump density. From extrapolates
$\tau_{S}=(1.6\pm0.1)~\mu\text{s}$ for the data from sample~\#1 at
$T=1.8$~K and $B_\mathrm{F}=5$~mT. Performing measurements in the
same experimental conditions for sample~\#2 we obtain
$\tau_{S}=(1.1\pm0.1)~\mu$s. In fact, the spin relaxation time of
donor-bound electrons is weakly dependent on the donor
concentration, which is not very surprising keeping in mind the
strong localizing potential of the fluorine donors in ZnSe and that
the highest donor concentration we examine is approximately
$1\times10^{18}$~cm$^{-3}$, i.e. the average distance between
neighboring donors is about 10 nm.

The evaluated spin relaxation time $\tau_{S}$ in dependence on the
magnetic field is shown in Fig.~\ref{fig:4}(a) for the range of weak
magnetic fields $B_\mathrm{F}<20$~mT. It is constant in this field
range at $\tau_{S}=1.6~\mu$s. Note in particular that it is also
constant below 5~mT where the electron spin polarization decreases
considerably due to the fluctuating nuclear magnetic field as can be
seen from the black line which shows the corresponding PR signal at
$f_{\text{m}}=75~\text{kHz}$. Furthermore, the results presented in
Figs.~\ref{fig:4}(a) and \ref{fig:4}(c) let us conclude that the
spin lifetime $T_{S}$ and, correspondingly, the spin relaxation time
$\tau_{S}$ do not depend on $B_{\text{F}}$ in the whole range from
zero up to 2.5~T. Note, that this is a rather unexpected result as
commonly the carrier spin relaxation time is sensitive to the
application of magnetic fields. Figure~\ref{fig:5} demonstrates
another surprising observation, namely that the spin relaxation time
does not depend on the temperature in the range from 1.8 to 45~K. We
will discuss in Sec.~\ref{sec:discussion} possible spin relaxation
mechanisms that can be responsible for this behavior.

Let us turn to the results recorded for CW pump, shown in
Fig.~\ref{fig:6}(a). Here the spin polarization of donor-bound
electrons is reduced in transverse magnetic field $B_{\text{V}}$,
showing the known behavior for the Hanle effect. The common
interpretation of the Hanle effect, when only an external magnetic
field acts on the electron spin, is as follows.  The external field
$B_{\text{V}}$ induces Larmor precession of the electron spins
around the magnetic field direction ($x$-axis). The frequency of
this precession is given by
$\Omega_{\text{L}}=\mu_{\text{B}}g_{\text{e}}B_{\text{V}}/\hbar$~\cite{OptOr}.
For relatively strong magnetic fields, for which the spin lifetime
$T_S$ is long compared to the time scale determined by
$\Omega_{\text{L}}$, the electron spins perform many revolutions
during their lifetime. This precession reduces the projection of the
spins on the initial direction ($z$-axis). The behavior is described
by a Lorentz curve (the Hanle curve, see Eq.~(50) in
Ref.~\cite{OptOr}).
\begin{equation}
\label{n12} S_{z}(B)=\frac{S_{0}}{1+(B/B_{1/2})^{2}},
\end{equation}
Here $B_{1/2}=\hbar/(\mu_{\text{B}}g_{\text{e}}T_{S})$ is the
characteristic field corresponding to the HWHM of
the Hanle curve.

For localized electrons in semiconductors, the hyperfine fields of
the nuclear spin fluctuations can also contribute to the Hanle
curve. In the limit of a long dwell time of the electrons on the
donors $\Omega_N \tau_{\text{c}}\gg 1$ the half width of the Hanle
curve will be controlled by the hyperfine field,
$B_{1/2}=\hbar/(\mu_{\text{B}}g_{\text{e}}T^*_{2})$, and the
respective time is the spin dephasing time $T^*_{2}$. Here
$\tau_{\text{c}}$ is the correlation time of an electron and a
donor. Using Eq. \eqref{n12} to fit the experimental data by
Fig.~\ref{fig:6}(a) we determine $B_{1/2}=0.95$~mT. In combination
with $|g_{\text{e}}|=1.13$ for the donor-bound electron we
obtain $T_{2}^{*}=10.5~\text{ns}$.

Alternatively, information on $T_{2}^{*}$  can be obtained from
the PR data shown in Fig.~\ref{fig:6}(b). The observed increase of
the electron spin polarization for higher magnetic fields can be
interpreted as a suppression of the influence of the nuclear spin
fluctuations by the external magnetic field. The width of the
observed dip of the PR curve can be used as a direct measure of
these fluctuations, which are characterized by an average hyperfine
nuclear field $B_{\text{N}}$~\cite{Greilich12,Petrov2008}. The
respective field dependence of the PR amplitude can be described by
\begin{equation}
\Theta_{\text{KR}}=\Theta_{0}\left[1-\frac{2/3}{1+(B/B_{\text{N}})^2}\right],
\label{e8a}
\end{equation}
where $\Theta_{\text{KR}}$ is the Kerr rotation angle, and
$\Theta_{0}$ is its value at zero magnetic field. The fit to the PR
data with Eq.~\eqref{e8a} shown in Fig.~\ref{fig:6}(b) yields
$B_{\text{N}}=1.9$~mT. From this value one calculates $T_{2}^{*}$ according to the following equation
\cite{Petrov2008}:
\begin{equation}
T_{2}^{*}=2\sqrt{3}\hbar/(\mu_{\text{B}}g_{\text{e}}B_{\text{N}})
\end{equation}
This calculation yields $T_{2}^{*}=18.3~\text{ns}$, which is in good agreement with the
dephasing time evaluated from the Hanle curve.

Additional arguments for the validity of our interpretation come
from the comparison of the RSA and PR signals measured under the
same experimental conditions, see Fig.~\ref{fig:2}(b). There the PR
amplitude (C) and the RSA peak amplitude at zero magnetic field (D)
exhibit a ratio of about 2:1. This ratio is in good agreement with
the conclusions of Ref.~\cite{Merkulov2002}. The Kerr rotation
signal is a direct measure of the spin polarization along the
direction of observation ($z$-axis). For weak or zero external
magnetic fields the components of the nuclear field $B_{\text{N}}$
oriented perpendicular to the direction of observation (i.e., the
$x$- and $y$-components) depolarize the electron spins, while the
components parallel or antiparallel to this axis ($z$-component) do
not alter the spin polarization. Due to the isotropic nature of the
nuclear fluctuations each of these three components of
$B_{\text{N}}$ has equal strength or probability, so that the spin
polarization of the electron interacting with $B_{\text{N}}$ is
reduced to one third of its maximum value at zero external magnetic
field~\cite{Merkulov2002, Petrov2008}.

\section{Discussion}
\label{sec:discussion}

In this section we compare the characteristic times of the electron
spin dynamics determined by the various techniques and discuss the
spin relaxation mechanisms which can describe the measured
properties of electron spins bound to fluorine donors in ZnSe.
Comparing the results of the different techniques we find that
$\tau_{S}{\gg}T_2^*$, i.e. the characteristic time determined from
the Hanle curve $T_{2}^{*}$ and the irreversible spin relaxation
time $\tau_{S}$ determined from the spin inertia method strongly
differ. This can be explained by broadening of the Hanle curve by
the nuclear spin fluctuations:  The strongly localized, donor-bound
electrons in the fluorine-doped ZnSe epilayers interact with the
nuclear hyperfine field of the same nuclei for a long time
($\tau_{\text{c}}\ge\hbar/(\mu_{\text{B}}g_{\text{e}}B_{\text{N}})$).
The resulting Larmor precession in the nuclear hyperfine field
broadens the Hanle curve, so that the spin lifetime $T_{S}$ obtained
from the Hanle measurement is limited by this reversible effect and
much shorter than the time for the irreversible spin relaxation
$\tau_{S}$ determined from the spin inertia method.

Every mechanism of irreversible spin relaxation can be interpreted
as the effect of temporally fluctuating magnetic fields on the
electron spin. Eq.~\eqref{n1} is valid in the case of short
correlation times of the random magnetic field
$\tau_{c}{\ll}\tau_{S}$, when dynamical averaging takes place. In a
strong magnetic field the relaxation times of the longitudinal and
transverse components $T_1$ and $T_2$ are different. The time $T_1$
describes the decay of the spin component along the magnetic field.
This time can depend considerably on the magnetic field, since the
spin-flip requires the transfer of the energy
$\mu_{\text{B}}g_{\text{e}}B$ to the lattice. On the contrary, the
time $T_2$ describes the decoherence time, which is not related to
energy transfer to the lattice. They become equal to each other
$T_1=T_2=\tau_S$  in a weak magnetic field~\cite{Abragam}. The spin
relaxation time $\tau_{S}$ that we determine with the spin inertia
method with the magnetic field applied in Faraday geometry is the
$T_{1}$ time.

For sufficiently strong longitudinal magnetic fields one would
expect a dependence of the spin relaxation time on the magnetic
field. However, we do not observe any dependence of $\tau_{S}$ on
the magnetic field from zero up to 2.5 T for the donor-bound
electrons and only small variations within the accuracy of our
method in the temperature range from 1.8 up to 45~K. This imposes
severe restrictions on the fluctuating magnetic fields, which can be
used to describe the spin relaxation process. Calculating the Zeeman
splitting of the electron states at an external magnetic field of
$B=2.5~$T we can deduce that the fluctuations of the random magnetic
field describing the underlying relaxation mechanism must have a
wide frequency range $\mu_{\text B}g_{\text
e}B/\hbar\approx(3~\text{ps})^{-1}$. Thus, the correlation time of
the corresponding fluctuating field must be shorter than 3 ps.

The following, almost instantaneous processes can be considered: (i)
scattering between free and donor-bound electrons (the exchange
interaction between the electrons is responsible for the electron
spin flip), (ii) jumping of electrons between different donors
(hyperfine and spin-orbit interaction), (iii) scattering of phonons
by donor-bound electrons (spin-orbit interaction), and  (iv)  charge
fluctuations in the environment of the donors (spin-orbit
interaction). All of them will be discussed in the following.

The process (i) is unlikely, because the localized states are
excited resonantly and the donor ionization process should depend on
temperature in the range from 30 to 50 K, which does no reflect the
experimental observations.

The process (ii) can be provided by two mechanisms: Electron spin
flip-flop transitions, which are induced by the scalar exchange
interaction between electrons on neighboring donors, and electron
jumps of donor-bound electrons to unoccupied donors. Calculations
according to Ref.~\cite{Kavokin2008} for the parameters of the
fluorine donor in ZnSe yield a jump time much longer than the
estimated 3 ps. Thus, we discard option (ii) as a possible
mechanism.

We also discard the process (iii), since we do not observe any
temperature dependence of the spin relaxation time $\tau_{S}$, which
we would expect for a phonon-mediated process.

The only possible mechanism left is the process (iv) - charge
fluctuations in the environment of the donors, which e.g. might
occur during the 1.5 ps duration of the laser pulse illumination. We
test this possibility by changing the pulsed pump beam to CW
excitation. However, the determined spin relaxation time still does
not depend on magnetic field or on temperature in the specified
range. According to this check we can exclude a direct influence of
the pulsed excitation on the spin relaxation. However, we cannot
completely disregard any illumination induced mechanism, as charge
fluctuations can be produced also by CW laser excitation in
combination with carrier recombination during tens of
picoseconds~\cite{Kim2014}. Still, there is no clear evidence for this
so that we suggest that there may be a different, new
mechanism, which determines the spin relaxation time in this system
with strong electron localization.

\section{Conclusions}
\label{sec:conclusions}

We have suggested a method based on the spin inertia effect to
measure the longitudinal spin relaxations time $T_1$ of carriers. It
exploits optical orientation of the carrier spins and their
polarization recovery in magnetic field in the Faraday geometry,
measured for different modulation frequencies of the laser helicity.
The validity of this method is demonstrated for electrons bound to
fluorine donors in ZnSe. An electron spin relaxation time of
$T_1=1.6$~$\mu$s is measured for sample \#1 in the temperature range
$1.8-45$~K. This time remains constant for magnetic fields varied
from zero to 2.5~T and depends only weakly on the donor
concentration. Measurements of the spin dephasing time
$T^*_2=8-33$~ns by the RSA and Hanle techniques, and comparison of
pulsed and continuous-wave excitation allow us to conclude that the
spin relaxation of the donor-bound electrons is caused by
perturbations that cover a broad spectral range. The question about
the origin of this perturbation has remained open so far and needs
further investigations.

The obvious advantage of the suggested polarization recovery
technique based on the spin inertia effect is that it is suitable
for measuring the longitudinal spin relaxation time $T_1$ in the
whole range of magnetic fields starting from zero field.
Contributions of different spin relaxation mechanisms may be
distinguished by their different onsets in the modulation frequency
dependence. This distinction is possible when the generated carrier
spin polarization is not fully destroyed by a faster relaxation
mechanism. A requirement for the suggested technique is the finite
optical orientation of carriers (at least of about few percent that
can be comfortably detected). The photoinduced carrier spin
polarization can be detected by various methods, e.g. by Kerr or
Faraday rotation or by the circular polarization degree of
photoluminescence. The main limitation of the technique comes from
the condition that the pump helicity modulation period shall be
tuned to a time shorter than the spin lifetime $T_S$. Therefore, the
technique can be applied well for measuring long relaxation times,
e.g. of resident carriers, but it is less suited for fast decaying
excitons, for example, whose typical recombination time is shorter
than a nanosecond, as it would require a modulation frequency
exceeding 1~GHz.

{\bf Acknowledgements} This work has been supported by the
Deut\-sche For\-schungs\-ge\-mein\-schaft via
Sonderforschungsbereich TRR 160, the Volkswagen Stiftung
(Project-No. 88360) and the Russian Science Foundation (Grant No.
14-42-00015). V.L. Korenev acknowledges financial support of the
Deut\-sche For\-schungs\-ge\-mein\-schaft within the Gerhard
Mercator professorship program.

\end{document}